\documentclass[a4paper,12pt]{article}
\pdfoutput=1
\usepackage[page,toc,titletoc,title]{appendix}
\usepackage{amsmath}
\usepackage{amssymb}
\usepackage[small]{caption}
\usepackage{cite}
\usepackage[top=1.2in, bottom=1.2in, left=1.1in, right=1.1in]{geometry}
\usepackage[multiple]{footmisc}
\usepackage{graphicx}
\usepackage[nottoc]{tocbibind}
\usepackage{xcolor}

\definecolor{mediumblue}{rgb}{0,0,0.8}
\usepackage{hyperref}
\hypersetup{
  linktocpage=true,
  colorlinks=true,
  citecolor=mediumblue,
  filecolor=mediumblue,
  linkcolor=mediumblue,
  urlcolor=mediumblue,
}
\newcommand{\mailref}[1]{E-mail: \href{mailto:#1}{#1}}

\graphicspath{{./}{./figures/}}
\numberwithin{equation}{section}
\allowdisplaybreaks%

\def\thefootnote{\fnsymbol{footnote}}

\newcommand{\abs}[1]{\left|{#1}\right|}
\newcommand{\GeV}{\mathrm{GeV}}
\newcommand{\sel}{\tilde{\ell}}
\newcommand{\la}{\lambda}
\newcommand{\br}[2]{\mathrm{BR}\left({#1}\to{#2}\right)}

\begin{document}

\begin{titlepage}
  \begin{flushright}
    \texttt{PNUTP-21-A13}, \texttt{CTPU-PTC-21-25}
  \end{flushright}

  \bigskip

  \begin{center}
    \bf\boldmath\Large
    Mixed modulus and anomaly mediation 
    \\[2mm]
    in light of the muon $g-2$ anomaly
  \end{center}

  \medskip

  \begin{center}
    Kwang~Sik~Jeong$^{a,}$\footnote{\mailref{ksjeong@pusan.ac.kr}},
    Junichiro~Kawamura$^{b,c,}$\footnote{\mailref{jkawa@ibs.re.kr}},
    and
    Chan~Beom~Park$^{b,}$\footnote{\mailref{cbpark@ibs.re.kr}}
  \end{center}

  \begin{center}
    \small
    $^a${\em Department of Physics, Pusan National University, Busan
      46241, Korea}\\[0.1cm]
    $^b${\em Center for Theoretical Physics of the Universe,
    Institute for Basic Science (IBS),\\
    Daejeon 34126, Korea}\\[0.1cm]
    $^c${\em Department of Physics, Keio University, Yokohama
      223--8522, Japan}
  \end{center}

  \medskip

  \begin{abstract}
    The new measurement of the anomalous magnetic moment of
    muon at the Fermilab Muon $g-2$ experiment has strengthened the
    significance of the discrepancy between the standard model
    prediction and the experimental observation from the BNL
    measurement.
    If new physics responsible for the muon $g-2$ anomaly is
    supersymmetric,   one should consider how to obtain light
    electroweakinos and sleptons in a systematic way.
   The gauge coupling unification allows a robust prediction of the
   gaugino masses, indicating that the electroweakinos can be much
   lighter than the gluino if anomaly-mediated supersymmetry breaking
   is sizable.
   As naturally leading to mixed modulus-anomaly mediation, the KKLT
   scenario is of particular interest and is found capable of
   explaining the muon $g-2$ anomaly in the parameter region where the
   lightest ordinary supersymmetric particle is a bino-like neutralino
   or slepton.
  \end{abstract}
\end{titlepage}

\renewcommand{\thefootnote}{\arabic{footnote}}
\setcounter{footnote}{0}

\setcounter{tocdepth}{2}
\noindent\rule{\textwidth}{0.3pt}\vspace{-0.4cm}\tableofcontents
\noindent\rule{\textwidth}{0.3pt}


\section{Introduction}

\noindent
Since the experimental observation of the discrepancy at the
Brookhaven National Laboratory (BNL)~\cite{Bennett:2004pv}, the anomalous
magnetic moment $g - 2$ of muon has served as a long-standing puzzle of
particle physics.
Recently, the Fermilab Muon $g-2$ collaboration has announced the new
measurement result~\cite{Abi:2021gix}, which has further strengthened the
significance of the BNL result on the muon $g-2$. Comparing to the
Standard Model (SM) prediction~\cite{
  czarnecki:2002nt, melnikov:2003xd, aoyama:2012wk, gnendiger:2013pva,
  kurz:2014wya, colangelo:2014qya, masjuan:2017tvw, Colangelo:2017fiz,
  davier:2017zfy, keshavarzi:2018mgv, hoferichter:2018kwz,
  colangelo:2018mtw, Aoyama:2019ryr, gerardin:2019vio,
  hoferichter:2019gzf, davier:2019can, bijnens:2019ghy,
  colangelo:2019uex, keshavarzi:2019abf, Blum:2019ugy, Aoyama:2020ynm
},
the combined BNL and Fermilab result amounts be a $4.2\sigma$ discrepancy.
The deviation from the SM prediction is
\begin{equation}
  \Delta a_\mu = a_\mu^\text{exp} - a_\mu^\text{SM} = (25.1 \pm 5.9)
  \times 10^{-10} ,
\end{equation}
where $a_\mu \equiv (g_\mu - 2) / 2$.
Although the recent lattice calculation for the hadronic vacuum
polarization contribution to the muon $g-2$ has turned out to be in
accord with the measured value~\cite{Borsanyi:2020mff},\footnote{
  In Refs.~\cite{Lehner:2020crt, Crivellin:2020zul,
    Keshavarzi:2020bfy, Malaescu:2020zuc},
  it has been claimed that shifting the hadronic vacuum
  polarization value of the SM to match the measured value of the muon
  $g-2$ would result in tension with the global fit to electroweak
  precision data.
}
we take this
opportunity to examine the possibility of new physics accounting for the
muon $g-2$ anomaly.

Among various possible models that can explain the muon $g-2$ anomaly,
we consider the supersymmetric (SUSY) model as it is one of the most
promising candidates.
Ever since the announcement of the Fermilab result has come out, the
SUSY interpretations for the muon $g-2$ anomaly have already been
revisited or renewed in many works~\cite{Crivellin:2021rbq,
  Endo:2021zal, Gu:2021mjd, VanBeekveld:2021tgn, Yin:2021mls,
  Wang:2021bcx, Abdughani:2021pdc, Ibe:2021cvf, Cox:2021gqq,
  Han:2021ify, Baum:2021qzx, Zhang:2021gun, Ahmed:2021htr,
  Yang:2021duj, Athron:2021iuf, Aboubrahim:2021rwz,
  Chakraborti:2021bmv, Baer:2021aax, Altmannshofer:2021hfu,
  Aboubrahim:2021phn, Chakraborti:2021squ, Zheng:2021wnu,
  Zhang:2021nzv
}.
In the Minimal Supersymmetric Standard Model (MSSM), the contributions
to the muon $g-2$ can be generated by bino, winos, Higgsinos, smuons,
and sneutrino.
For sparticle masses of order $M_\text{SUSY}$, the leading SUSY
contributions have the generic behavior of
\begin{equation}
  \Delta a_\mu^\text{SUSY} \propto \frac{m_\mu^2 \, \mu
    M_a}{M_\text{SUSY}^4} \tan\beta ,
\end{equation}
where $\mu$ is the Higgsino mass, $M_a$ is the gaugino mass, and
$\tan\beta$ is the ratio of the Higgs vacuum expectation values~\cite{Stockinger:2006zn}.
Their relative contributions can differ by the
sparticle mass spectrum.
For example,   if the Higgsinos are very heavy,
only the bino-smuon loop contribution can become sizable.
Whereas the sfermion masses are highly model dependent,
the gaugino masses show a robust pattern
owing to the gauge coupling unification
at the grand unified theory (GUT) scale,  $M_{\rm GUT}$~\cite{Choi:2007ka}.
%
For instance,  in gravity mediation with universal gaugino masses at $M_{\rm GUT}$~\cite{Nilles:1982ik}
or in gauge mediation with messengers forming a GUT multiplet~\cite{Dine:1981za,Dimopoulos:1981au,Dine:1981gu},
the ratios of low energy gaugino masses read
\begin{equation}
  M_1 : M_2 : M_3 \simeq 1 : 2 : 6
\end{equation}
at the TeV scale,
regardless of the details of the model.
Anomaly mediation~\cite{Randall:1998uk,Giudice:1998xp},
which always exists in supergravity,  modifies
the above gaugino mass relation depending on its relative strength,
and may make the wino and/or bino much lighter than the gluino
as is required to explain the muon $g-2$ anomaly.
A natural framework for sizable anomaly mediation is provided by the KKLT string compactification~\cite{Kachru:2003aw}.
A remarkable feature of the KKLT moduli stabilization is that
the parameters of SUSY breaking are, in principle,    controlled by discrete numbers,
such as the winding number of D-branes, the number of fluxes that generate moduli potential, and so on.
%

In this article,
we point out that
mixed modulus-anomaly mediation~\cite{Choi:2004sx, Choi:2005ge, Endo:2005uy,
Choi:2005uz, Falkowski:2005ck}, which is realized in the KKLT setup,
can accommodate light electroweakinos (EWinos) and sleptons to explain
the muon $g-2$ anomaly and heavy colored sparticles to evade the lower
limits from the LHC\@.
To obtain the suitable sparticle mass spectra for the muon $g-2$, we
need to consider a generalized KKLT setup beyond the minimal one,   as
described in Sec.~\ref{sec:kklt}.
By imposing various conditions such as the Higgs boson
properties
and the vacuum stability of the scalar potential, we perform numerical
analysis in the parameter space to check the feasibility of the model.
We present our analysis result in Sec.~\ref{sec:g-2}.
Although our result is mostly safe from the lower limits set by the
search results on colored sparticles at the LHC, the bounds from the
searches for the chargino-neutralino and the slepton pair productions
may exclude the parameter points of light sleptons and gauginos.
We also find that,
in a large part of the parameter space for the muon $g-2$,
a slepton becomes lighter than the lightest neutralino.
In this case, we should consider alternative scenarios such as a
Peccei-Quinn (PQ) symmetric extension where the axino is the lightest
sparticle or R-parity violating (RPV) interactions to make the
lightest ordinary SUSY particle (LOSP), the lightest sparticle among
the MSSM sparticles, unstable.
The LHC limits and the phenomenological
scenarios with light sleptons are discussed in Sec.~\ref{sec:lhc}.
We summarize our study in the last section.
For reference,   we list the
benchmark sparticle mass spectra in Appendix~\ref{sec:benchmark}.

\section{\label{sec:kklt}\boldmath Mixed modulus-anomaly mediation}

\noindent
The sparticle mass spectrum crucially depends on how SUSY breaking in a hidden sector is
transmitted to the visible sector.
To be consistent with the experimental constraints, SUSY breaking
mediation should preserve flavor and CP symmetry with good accuracy
unless it makes the sparticles very heavy above $100$~TeV.
Indeed, many mediation schemes such as gravity mediation,
dilaton/moduli mediation, gaugino mediation, gauge mediation, anomaly
mediation, and their mixtures conserve flavor and CP symmetry and lead
to various patterns of sparticle spectra.
Among the sparticles, the gauginos are known to have a robust pattern of
masses under the condition of gauge coupling
unification~\cite{Choi:2007ka}.
The gaugino masses in mixed modulus-anomaly mediation are written as
\begin{equation}
M_a = M_0 \left( 1 + \frac{b_a g^2_{\rm GUT} }{4} \alpha \right),
\end{equation}
at the scale just below $M_{\rm GUT}$,
where the gauge coupling constants have
the common value,   $g^2_a(M_{\mathrm{GUT}}) = g^2_{\rm GUT}$.
Here, $b_a=(33/5,\,1,\,-3)$ are the coefficients of the one-loop beta functions
at TeV,   and the $\alpha$ parameter represents the relative strength of
anomaly mediation:
\begin{equation}
\alpha \equiv \frac{m_{3/2}}{4\pi^2  M_0},
\end{equation}
with $m_{3/2}$ being the gravitino mass.
Note that anomaly mediation is a model-independent supergravity effect
proportional to the gravitino mass,
but it alone suffers from the tachyonic
slepton problem.
Because the combination $M_a/g^2_a$ is renormalization group (RG) invariant
at one loop, the gaugino masses at the TeV scale are found to
approximately obey the following relation:
\begin{equation}
\label{Ma-TeV}
M_1 : M_2 : M_3 \simeq (1 + 0.83 \alpha) :
(2 + 0.25 \alpha) : (6 -2.25 \alpha),
\end{equation}
where we have taken $g^2_{\rm GUT} =0.5$ and used the ratios of the
gauge couplings $g^2_1 : g^2_2: g^2_3 \simeq 1:2:6$ at the TeV scale.

As the first explicit realization of a de Sitter (dS) vacuum with all
string moduli stabilized, the KKLT mechanism~\cite{Kachru:2003aw} provides an interesting
framework to realize mixed modulus-anomaly mediation with $\alpha$ of
order unity.
In the minimal KKLT setup,  $\alpha$ is a positive rational
number~\cite{Choi:2004sx,Choi:2005ge}.
In the literature,
the phenomenology of positive $\alpha$ has been studied intensively
particularly to resolve the fine-tuning problem to realize the electroweak (EW) symmetry breaking scale~\cite{Kitano:2005wc,Lebedev:2005ge,Choi:2006xb,Abe:2014kla,Baer:2016hfa,Kawamura:2017qey,Jeong:2020wum}.
However, the muon $g-2$ anomaly is hardly explained in the cases with
positive $\alpha$ due to relatively heavy winos and bino unless the
other contribution exists~\cite{Choi:2006xb, Du:2018pko}.
One can generalize the KKLT setup to obtain a negative $\alpha$ to get
a desired mass hierarchy in the gaugino spectrum.
For a concrete discussion, let us consider a model where the effective
moduli superpotential is  given by\footnote{
  We take the reduced Planck mass unit, $M_{Pl}=1$, unless stated otherwise.
}
\begin{equation}
W = A_0 e^{-4\pi^2 \ell_0 S_0}
- A_1 e^{-4\pi^2(k_1 T + \ell_1 S_0)},
\end{equation}
with $A_0$ and $A_1$ being constants of order unity,
and the visible sector gauge kinetic function is written as
\begin{equation}
\label{fa}
f_a = k T  + \ell S_0,
\end{equation}
after integrating out the heavy dilation $S$ and
complex structure moduli
fixed by fluxes at $S=S_0$.
Here, $\ell_0/\ell$,  $\ell_1/\ell$, and $k_1/k$ are rational numbers determined by topological or
group theoretical data of the underlying string compactification,
as can be dictated from the periodicities of ${\rm Im}(T)$ and ${\rm Im}(S)$.
A supersymmetric minimum is developed by the above superpotential
and is lifted to a dS vacuum by adding a SUSY breaking uplifting
potential that originates from a brane-localized source located at the
IR end
of the warped throat.
The uplifting potential is given by
\begin{equation}
V_{\rm lift}
= \frac{P e^{2K_0/3}}{(T+T^*)^{n_P} },
\end{equation}
for a rational number $n_P$.
$P$ is a positive constant fixed by the condition of vanishing cosmological constant.
Here, the modulus K\"ahler potential generally reads
\begin{equation}
K_0= -n_0 \ln(T+T^*),
\end{equation}
for a positive rational number $n_0$.
An extra-dimensional interpretation of the uplifting procedure
is possible for $n_P\geq 0$, because otherwise
the uplifting sector couples more strongly for a larger value of $T$~\cite{Aad:2015zhl}.
In the above model, one finds $\alpha$ to be~\cite{Choi:2006xb}
\begin{equation}
\alpha
 \equiv
 \frac{m_{3/2}}{4\pi^2 M_0}
\simeq
\frac{2k_1}{k}
\left( 1 + \frac{3n_P}{2n_0} \right)^{-1},
\end{equation}
from the observations
that gauge coupling unification requires
$g^{-2}_{\rm GUT} = {\rm Re}(f_a)   \simeq 2$,
and that $T$ is stabilized at $k_1 T \simeq (\ell_0 -\ell_1) S_0$ with
$S_0$ written in terms of the gravitino mass as
$4\pi^2 \ell_0 {\rm Re}(S_0) \simeq \ln(M_{Pl}/m_{3/2})$.
Note that $M_0$ is given by $M_0=F^T\partial_T \ln {\rm Re}(f_a)$ with $F^T$ being the
modulus $F$-term.
A negative $\alpha$ is therefore obtained if either $k_1$ or $k$ is
negative.
See,  for example,  Ref.~\cite{Abe:2005rx} for more discussion on the case with
$k_1<0$ and $k>0$.


Let us continue to examine the sfermion masses, which generally
possess a stronger model dependence compared to the gaugino masses.
In the mixed modulus-anomaly mediation under consideration, the
sfermion masses are determined by the modulus dependence of the matter
K\"ahler potential:
\begin{equation}
K = -n_0 \ln(T+T^*) + \frac{ \Phi^*_i \Phi_i }{(T+T^*)^{n_i} }.
\end{equation}
We have taken into account a simple case where the matter K\"ahler
metric is not affected by the involved dilaton-modulus mixing.
Here,   the modular weight $n_i$ is a rational number of order unity fixed by
the location of the matter in extra dimensions.
The mixed mediation in the KKLT preserves CP and flavor symmetries respectively due to the axionic
shift symmetry associated with $T$ and flavor-universal modular weights.
The pure modulus-mediated (MM) trilinear $A$-parameters and soft
scalar masses are found to be
\begin{align}
A_{ijk}|_{\rm MM} &= - (a_i +a_j + a_k) M_0,
\nonumber \\
m^2_i|_{\rm MM} &= c_i M^2_0,
\end{align}
at $M_{\rm GUT}$,  where $a_i$ and $c_i$ are given by
\begin{align}
a_i &= \left(  \frac{n_0}{3} - n_i \right)
\left(
1 + \frac{k_1}{k} \frac{\ell}{\ell_0-\ell_1}
\right),
\nonumber \\
c_i &=
\left(
1 + \frac{k_1}{k} \frac{\ell}{\ell_0-\ell_1}
\right)
a_i.
%
\end{align}
Note that $a_i$ and $c_i$ are rational numbers of either sign depending on the choice
of the associated discrete numbers.

It is worth noting that an anomalous U$(1)_A$ gauge symmetry allows
the $\alpha$ parameter and the
effective
modular weights to have various values
in a much wider range if $T$ transforms non-linearly to
implement the Green-Schwarz (GS) anomaly cancellation mechanism~\cite{Green:1984sg}.
Let us consider a simple case where the modulus-induced Fayet–Iliopoulos term is canceled by
a single U$(1)_A$ charged but SM singlet matter field $X$.
Integrating out  the heavy U$(1)_A$ gauge superfield,  whose longitudinal component
comes mostly from $X$,
one can construct  the  low energy effective theory of a light modulus $T$,  which is mainly the GS modulus,
and light matter fields.
The  K\"ahler potential reads~\cite{Choi:2006bh}
\begin{equation}
K_{\rm eff} = -n_0 \ln(T+T^*) + \frac{\Phi^*_i \Phi_i}{(T+T^*)^{n^{\rm eff}_i}},
\end{equation}
with the effective modular weight given by
\begin{equation}
n^{{\rm eff}}_i   \simeq   n_i + ( 1-n_X ) \frac{q_i}{q_X},
\end{equation}
where $n_\alpha$ and $q_\alpha$ are
the modular weight and U$(1)_A$ charge of the corresponding field, respectively.
From the fact that the superpotential is a holomorphic function of the U$(1)_A$ invariant combination
of the GS modulus and $X$,  the effective superpotential is found to be
\begin{equation}
 W_{\rm eff} =   A_0 e^{-4\pi^2 \ell_0 S_0}
- A_1 e^{-4\pi^2(k_1 T +k_H + \ell_1 S_0)}.
\end{equation}
for the constants $A_0$ and $A_1$ of order unity.
Here, $k_H$ is a constant of order unity fixed by the U$(1)_A$ invariance.
It is then straightforward to see that $\alpha$ is given by
\begin{equation}
\alpha  \simeq
\left(
1 - \frac{4\pi^2}{\ln(M_{Pl}/m_{3/2}) } k_H
\right)
\times \frac{2k_1}{k} \left( 1+ \frac{3n_P}{2 n_0} \right)^{-1}.
\end{equation}
The above shows that a positive $k_H$  can  flip the sign of  $\alpha$,
implying that U$(1)_A$ not only enlarges the possible range of modular weights,
but also makes it possible to achieve a negative $\alpha$ in the moduli stabilization with $k k_1>0$.
%
Meanwhile,   the holomorphic Yukawa coupling changes as
\begin{equation}
\lambda_{ijk} \to \lambda_{ijk} \epsilon^{-(q_i+q_j + q_k)/q_X},
\end{equation}
because it arises from the superpotential,  $X^{n_{ijk}} \Phi_i \Phi_j \Phi_k$,  where
$n_{ijk} =-(q_i + q_j + q_k)/q_X$ should be a non-negative integer.
$\epsilon\sim 0.1$ represents the VEV of
$X$ relative to $M_{Pl}$.
It is clear that a
large Yukawa coupling $y_{ijk}$ of order unity apparently constrains
the U$(1)_A$ charges to be $n_{ijk} = 0$ or $1$.\footnote{
In the case where $n_X=1$, one can assign flavor-dependent U$(1)_A$
charges, for which U$(1)_A$ can account for the hierarchical Yukawa
couplings via the Froggatt-Nielsen mechanism~\cite{Froggatt:1978nt}.
}

To summarize,  the generalized KKLT setup leads to mixed
modulus-anomaly mediation where the sparticle masses are determined by
three types of dimensionless parameters:
\begin{equation}
\alpha, \quad a_i,  \quad c_i,
\end{equation}
with $c_i \propto a_i$.
The parameters can take various values of order unity with either
sign, leading to a variety of sparticle mass spectra.
The overall size of sparticle masses is fixed by $M_0$.
While achieving gauge coupling unification, the gauginos show a robust
mass relation given by Eq.~(\ref{Ma-TeV}) at the TeV scale, and
interestingly, they can have a large mass hierarchy for a negative
$\alpha$.
\begin{figure}[tb!]
  \begin{center}
    \includegraphics[width=0.5\textwidth]{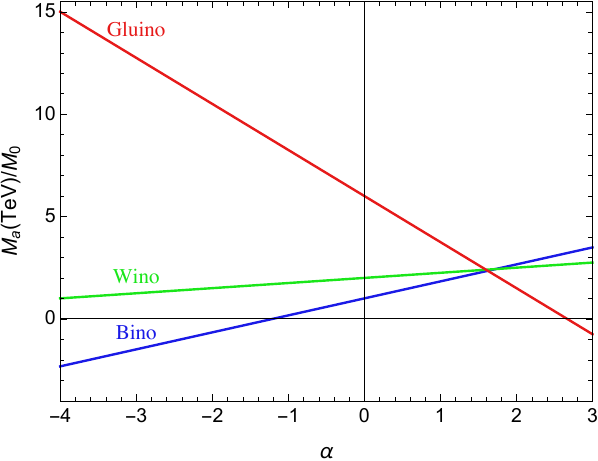}
  \end{center}
  \caption{\label{fig:gaugino_mass}
    Gaugino masses
    at TeV as functions of the $\alpha$
    parameter.  The masses have been normalized
    by $M_0$,  which is the pure modulus-mediated contribution
    at $M_{\rm GUT}$.
    The red,  green,  and blue  colored lines correspond to the
    gluino, wino, and bino masses,  respectively.}
\end{figure}
In Fig.~\ref{fig:gaugino_mass}, we display the gaugino masses as
functions of the $\alpha$ parameter, including negative values.
It shows that gluinos can become much heavier than the others
for a largely negative $\alpha$.
As discussed in the next sections, this feature is essential for
evading the collider bounds while explaining the muon $g-2$ anomaly
with light EWinos.
Meanwhile, the sfermion soft parameters just below the
unification scale are given by\footnote{
We employ the convention in the SUSY Les Houches Accord (SLHA)
format~\cite{Skands:2003cj} with $M_3$ being positive for $\alpha
\lesssim 2.5$.
For example, the parameters in Ref.~\cite{Choi:2006xb} can be
obtained by redefining $M_a \to -M_a$, $A_{ijk} \to -A_{ijk}$ and $\mu
\to -\mu$.
It corresponds to the field redefinitions:
$\la_a \to i \la_a$, $\psi_i \to i \psi_i$ and $\phi_i \to -\phi_i$,
where $\la_a$, $\psi_i$ and $\phi_i$ are gauginos, Weyl fermions, and scalars,
respectively. The other coupling constants are unchanged under the field redefinition.
The signs of the gaugino mass terms in the SLHA format
are opposite from those in Ref.~\cite{Choi:2006xb}.
Here, $A_{ijk}$ is a trilinear coupling divided by a Yukawa coupling
constant $y_{ijk}$.
}
\begin{align}
 \frac{A_{ijk}}{M_0}
&=
-(a_i + a_j + a_k)
- \frac{\alpha}{4} (\gamma_i + \gamma_j  + \gamma_k),
\nonumber \\
\frac{m^2_i}{M^2_0}
&=
c_i
+ \Big(
\sum_{jk} (a_i +a_j + a_k)  |y_{ijk}|^2
- 4 k \sum_a g^2_a C_a(\Phi_i)
\Big)  \frac{\alpha}{4}
+   \dot\gamma_i
\left( \frac{\alpha}{4} \right)^2,
\end{align}
for $f_a = k T + \Delta f_a$
with $\Delta f_a$ depending on other moduli of the model.
The anomalous dimension $\gamma_i$ is given by
\begin{equation}
  16\pi^2 \gamma_i = \frac{1}{2} y^{imn}y_{imn}-2g_a^2 C_a(\Phi_i) ,
\end{equation}
and $\dot\gamma = 8\pi^2 d\gamma_i/d \ln Q$
with $Q$ being the RG scale.
$C_a(\Phi_i)$ is the quadratic Casimir invariant of $\Phi_i$.

\section{\label{sec:g-2}\boldmath The muon $g-2$ anomaly}

\noindent
We are now in a position to examine the possibility of explaining the
muon $g-2$ anomaly in mixed modulus-anomaly mediation
realized in the generalized KKLT setup.
Depending on sparticle mass spectrum,  various different SUSY
contributions can enhance (or reduce) the muon $g-2$.
For a recent review on the SUSY contributions in light of the muon $g-2$ anomaly,
we refer the reader to Ref.~\cite{Athron:2021iuf} and the references
therein.
  In the MSSM, the most important contributions to the muon $g-2$
  arise from the Higgsino-wino-smuon (HWL) and bino-smuon (BLR) loop
  diagrams, which are given respectively by
\begin{align}
 \Delta a_\mu^{\mathrm{HWL}}=&\ \frac{g_2^2}{8\pi^2}
                                \frac{m_\mu^2 M_2}{m_{\tilde{\mu}_L}^4}
            \mu \tan\beta
              \left[F_a\left(\frac{M_2^2}{m_{\tilde{\mu}_L}^2},
                                   \frac{\mu^2}{m_{\tilde{\mu}_L}^2}\right)
            - \frac{1}{2} F_b\left(\frac{M_2^2}{m_{\tilde{\mu}_L}^2},
                                   \frac{\mu^2}{m_{\tilde{\mu}_L}^2} \right)\right], \\
 \Delta a_\mu^{\mathrm{BLR}} =&\ \frac{g_1^2}{8\pi^2} \frac{m_\mu^2}{M_1^3}
                                \mu \tan\beta F_b\left(\frac{m_{\tilde{\mu}_L}^2}{M_1^2},
                                   \frac{m_{\tilde{\mu}_R}^2}{M_1^2}
                                \right) .
\end{align}
The expressions for the loop functions $F_a$ and $F_b$ can be found in
e.g. Ref.~\cite{Athron:2021iuf}.
As will be discussed shortly,
the Higgsinos,  whose mass is tied to the up-type Higgs soft mass
under the condition of EW symmetry breaking,
are relatively heavy compared to other particles
relevant to the muon $g-2$.
In such a case,   the SUSY contribution to the muon $g-2$
mostly comes from the BLR one
because
the $\Delta a_\mu^{\mathrm{HWL}}$ is suppressed by large $\mu$.
For the same reason, the other loop effects involving the Higgsinos are subdominant.
Assuming that one of the smuons is significantly lighter than the other one,
the SUSY contributions to the muon $g-2$ are approximately given by
\begin{align}
    \label{eq:a_mu_BLR}
  \Delta a_\mu^\text{SUSY}
  &\approx \frac{g_1^2}{8\pi^2} \frac{m_\mu^2 \, \mu}{m_{\tilde\ell_+}^2 M_1}
    \tan\beta \times F_\mathrm{B}\left( \frac{m_{\tilde\ell_-}^2}{M_1^2} \right)
\\ \nonumber
  &\simeq
    2.5 \times 10^{-9}
    \Bigg(\frac{500~{\rm GeV}}{m_{{\tilde\ell}_+}}\Bigg)^2
    \Bigg(\frac{250~{\rm GeV}}{M_1}\Bigg)
    \Bigg(\frac{\mu}{2~{\rm TeV}}\Bigg)
    \Bigg(\frac{\tan\beta}{25}\Bigg)
    \Bigg(\frac{F_{\mathrm{B}}(m_{\tilde\ell_-}^2 /
    M_1^2)}{1/6}\Bigg)
    ,
\end{align}
where
\begin{align}
  F_B(x) = \frac{-1+x^2-2x\ln x}{2(x-1)^3}
\end{align}
is the loop function.
Here, $m_{\tilde\ell_- \,(\tilde\ell_+)}$ is the lighter (heavier)
smuon mass.
The
expression is valid as long as the SUSY contributions
to the muon $g-2$ are dominated by the BLR contribution and
the bino is much lighter than the heavier smuon.
It shows that the sign of the Higgsino mass $\mu$ and the
bino mass $M_1$ must be matched to have a positive contribution to
$\Delta a_\mu^\text{SUSY}$.

As we have seen in Sec.~\ref{sec:kklt},
the sparticle mass spectrum in mixed modulus-anomaly mediation
is governed by the three types of dimensionless parameters, $\alpha$, $a_i$,
$c_i$,  as well as
$M_0$.
Here we fix the $M_0$ value  by requiring that
the lightest CP-even Higgs boson,  whose properties approach to
those of the SM Higgs boson in the decoupling limit,
should have mass,  $m_h\simeq 125$~GeV,
to be compatible with the observation~\cite{Aad:2015zhl}.
For the sake of simplicity,
we take
\begin{equation}
a_i = c_i,
\end{equation}
which corresponds to the case
where the visible gauge kinetic function depends only on $T$,
i.e.~the case with $\ell=0$ in (\ref{fa}),  as in the minimal KKLT~\cite{Choi:2005ge}.
%
Furthermore,
motivated by  the flavor constraints and the unification of gauge couplings,
we assume that the modular weights respect the flavor universality and follow
the SU$(5)$ GUT relations for quarks and leptons,
\begin{equation}
 c_5    \equiv c_L = c_D, \quad
 c_{10}  \equiv c_Q = c_U = c_E.
\end{equation}
In our analysis,  therefore,
the input parameters of the model are given as follows:
\begin{equation}
 \alpha,\quad \tan\beta, \quad \mathrm{sgn}(\mu), \quad
 c_5,    \quad
 c_{10},  \quad
 c_{H_u},\quad c_{H_d}.
\end{equation}
Note  that the size of $\mu$ is determined by
the condition of EW symmetry breaking,
and  we take both signs of $\mu$ because $M_1$
can have either sign depending on the value of $\alpha$.
At  low energy scales,
the mass splittings of squarks and sleptons are induced by the RG
effects involved with the gauginos,  and anomaly-mediated
contributions.

To investigate the parameter space of mixed modulus-anomaly
mediation, 
we have added the boundary conditions of mixed mediation to the
\texttt{SOFTSUSY} program~\cite{Allanach:2001kg} and
calculated the sparticle and Higgs mass spectra using the program.
We require that the SM-like Higgs boson mass calculated with
\texttt{SOFTSUSY} is within $125.10 \pm 0.01$~GeV.
Then, for each parameter point,   we obtain the SUSY contributions to
$\Delta a_\mu$ by using \texttt{GM2Calc}, which can
compute the muon $g-2$ up to two-loop corrections~\cite{Athron:2015rva}.

Before looking into our analysis results, we should consider
some theoretical constraints.
The sleptons are required to be light to explain the muon $g-2$
anomaly, and the lightest stau can become tachyonic due to the large
values of $|\mu| \gtrsim 2$~TeV leading to a large mixing angle.
In our analysis, we discarded the parameter points with any tachyonic
sfermion, including sleptons, flagged by \texttt{SOFTSUSY}.
Furthermore, we should avoid the possibility of having a dangerous
charge-breaking minimum in the scalar potential deeper than the EW vacuum.
As it is not checked by \texttt{SOFTSUSY}, we impose the
vacuum stability condition given in Ref.~\cite{Endo:2013lva}: for
\begin{align}
  \label{eq:eta_l}
  \tilde{\eta}_\ell =
  &\  \abs{m^2_{\sel_{LR}}}\times
\left[ 101~\GeV\left( \sqrt{m_{\sel_L} m_{\sel_R}} + m_{\sel_L}
 + 1.03 m_{\sel_R} \right) \right.  \\ \nonumber
  &\ \left.  - 2.27\times 10^{4}~\GeV^2 +
  \frac{2.97\times10^6 \GeV^3}{m_{\sel_L}+m_{\sel_R}}
    - 1.14\times 10^8~\GeV^4 \left(\frac{1}{m_{\sel_L}^2}
    + \frac{0.983}{m_{\sel_R}^2} \right) \right]^{-1} ,
\end{align}
we require
\begin{equation}
  \tilde \eta_\ell < \eta_\ell ,
\end{equation}
where $\ell = \mu$, $\tau$.
In Eq.~(\ref{eq:eta_l}),
$m^2_{\sel_L}$, $m^2_{\sel_R}$, and $m^2_{\sel_{LR}}$
are the diagonal element of left-handed smuon,
that of right-handed smuon, and the off-diagonal element in the smuon
mass squared matrix, respectively.
Ignoring the small $\tan\beta$ dependence, we take $\eta_\tau$
($\eta_\mu$) $= 0.94$ ($0.88$)~\cite{Kitahara:2013lfa}.

\begin{figure}[tb!]
  \begin{center}
    \includegraphics[width=0.48\textwidth]{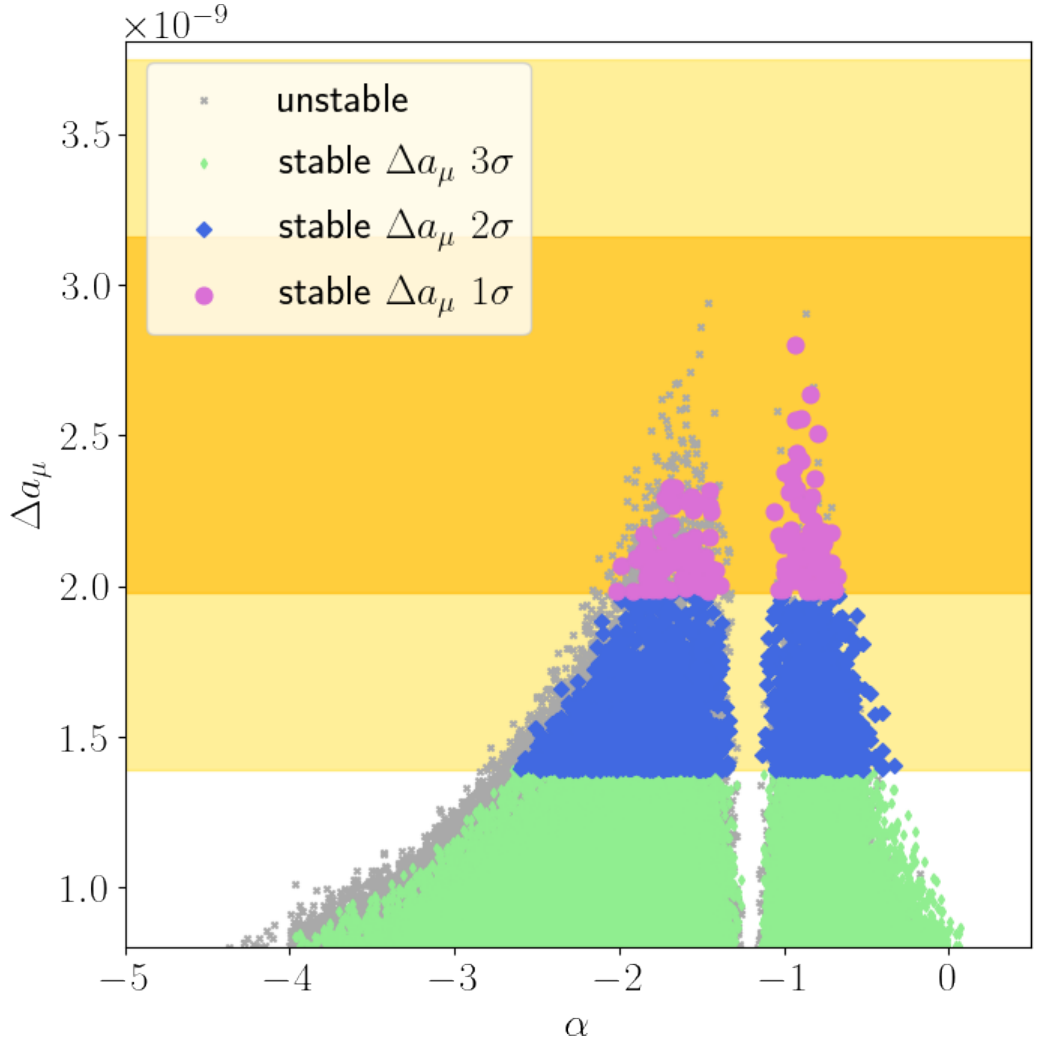}
    \includegraphics[width=0.48\textwidth]{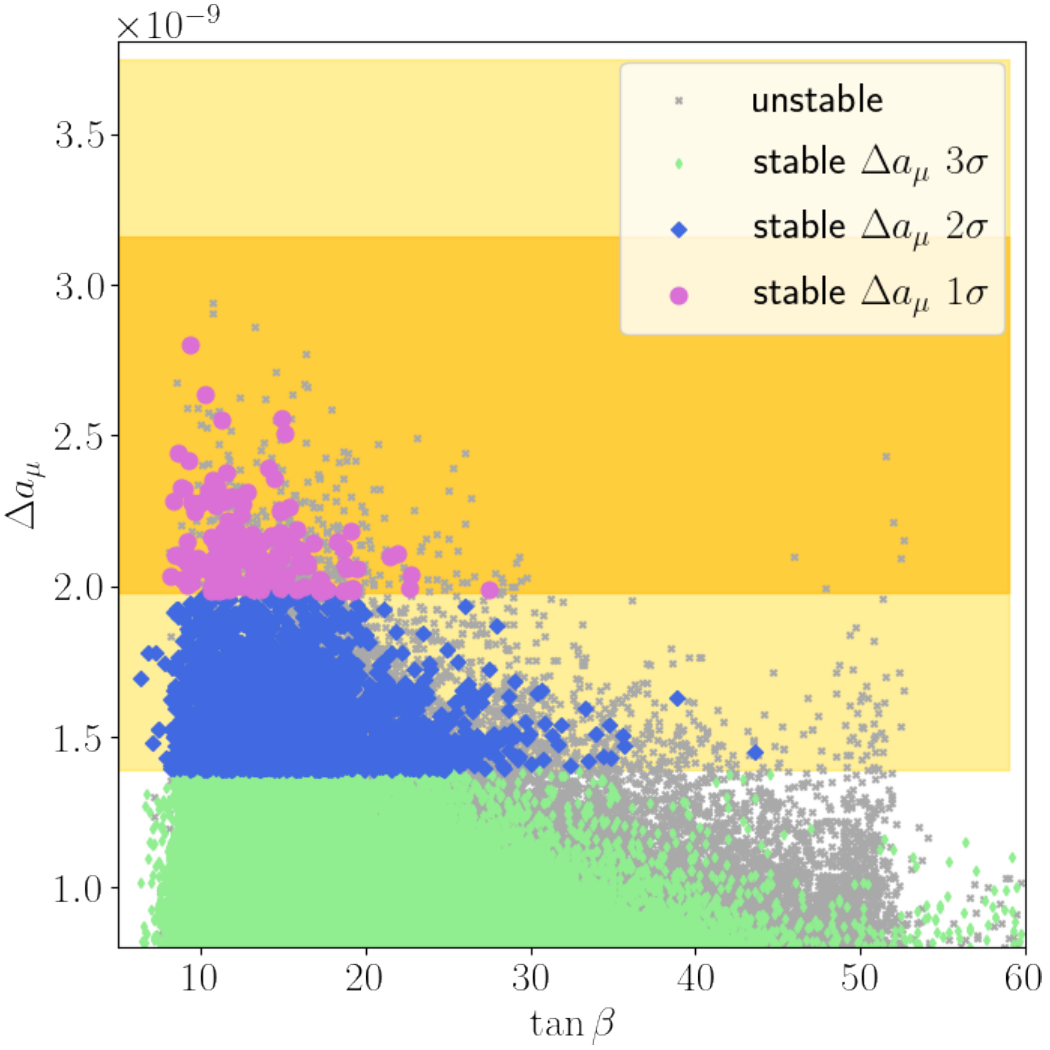}
  \end{center}
  \caption{\label{fig-scat_tbalp}
    Parameter points of $\alpha$ (left) and $\tan\beta$ (right)
    compatible with the muon $g-2$ anomaly. The orange and yellow
    bands correspond respectively to the $1\sigma$ and $2\sigma$
    ranges of the measured $\Delta a_\mu$ value, and accordingly, the
    parameters points are colored in magenta ($1\sigma$), blue
    ($2\sigma$), and green ($3\sigma$).
    Points colored in gray violate the vacuum stability condition.}
\end{figure}
\begin{figure}[tb!]
  \begin{center}
    \includegraphics[width=0.48\textwidth]{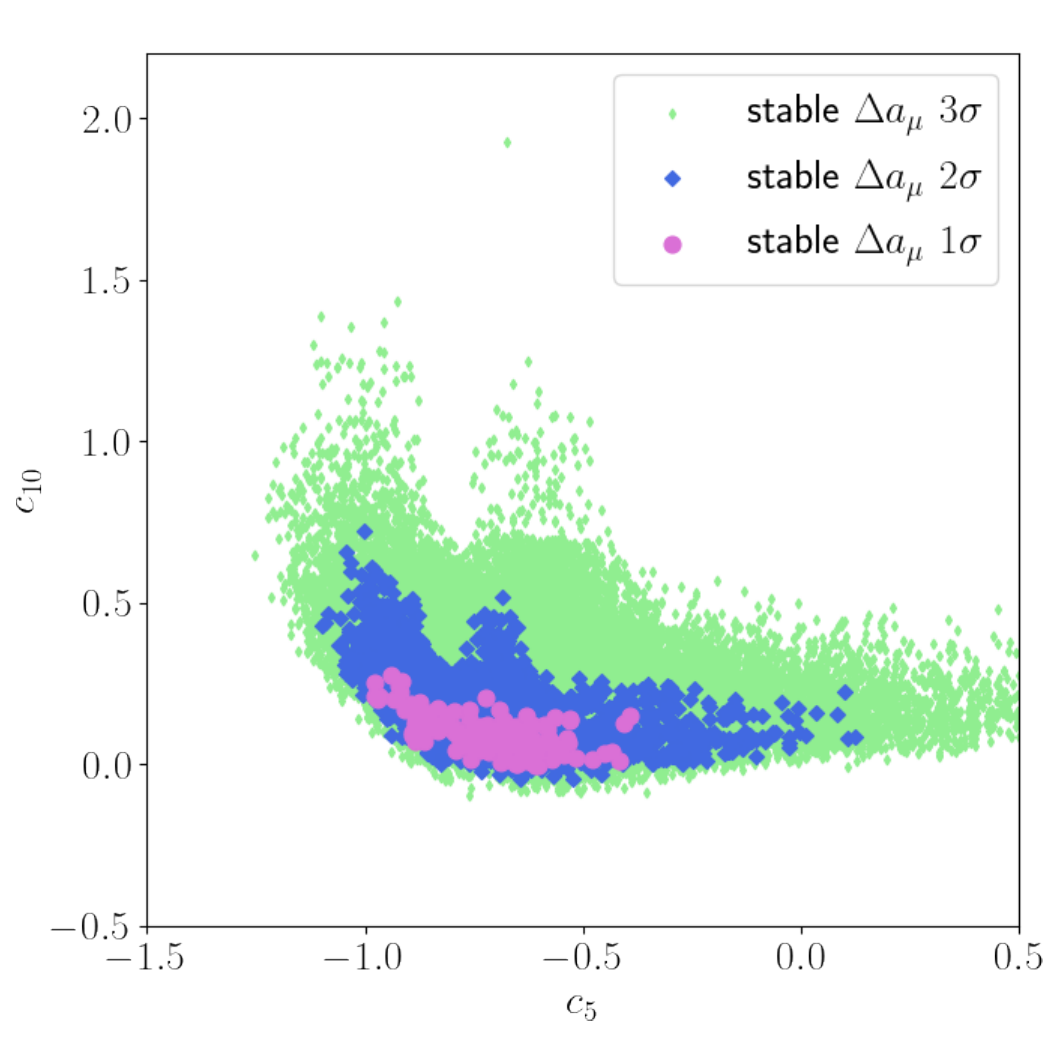}
    \includegraphics[width=0.48\textwidth]{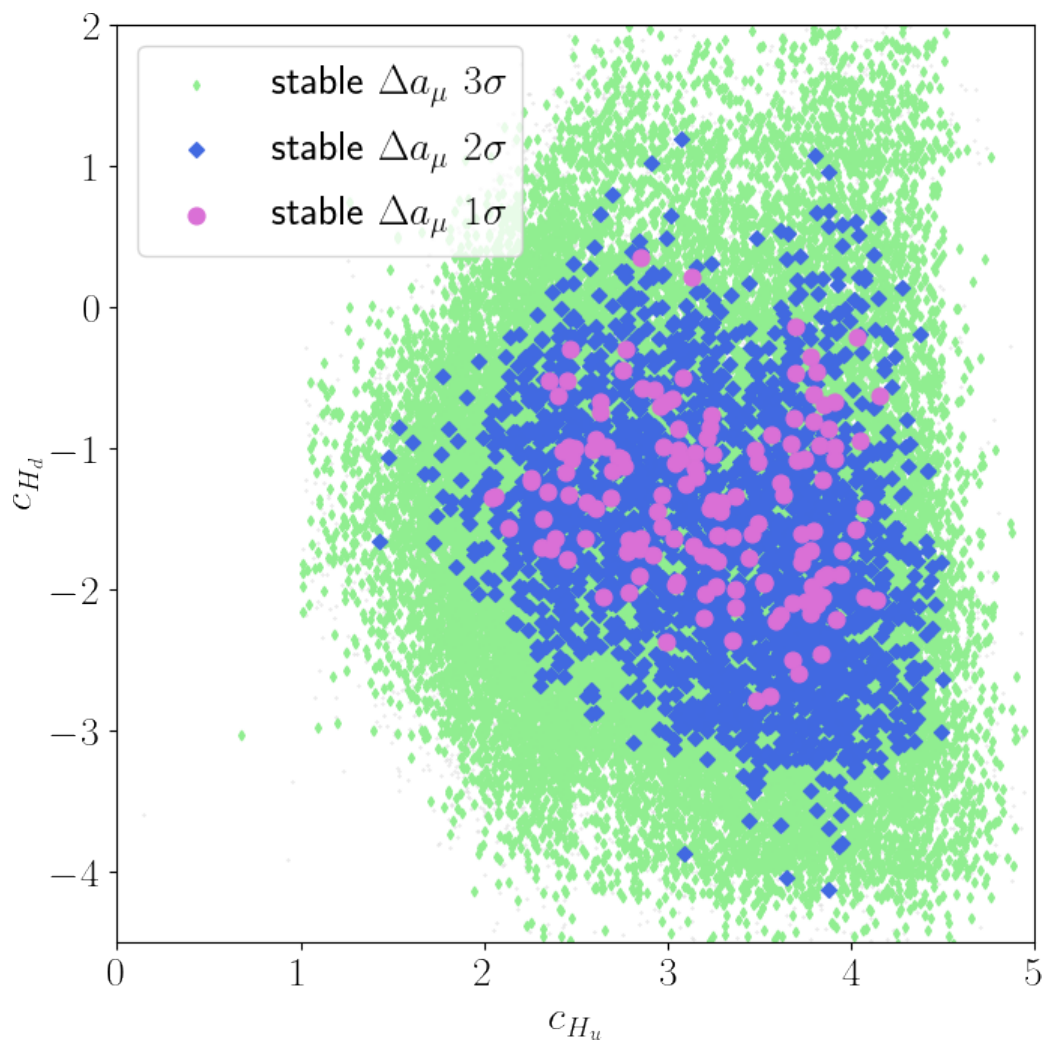}
  \end{center}
  \caption{\label{fig-scat_cs}
    Parameter points in the space of ($c_5$, $c_{10}$) (left) and ($c_{H_u}$,
    $c_{H_d}$) (right). The points colored in magenta, blue, and green
    can explain the muon $g-2$ anomaly within $1\sigma$, $2\sigma$,
    and $3\sigma$, respectively.}
\end{figure}

\begin{figure}[tb!]
   \centering
    \includegraphics[width=0.5\textwidth]{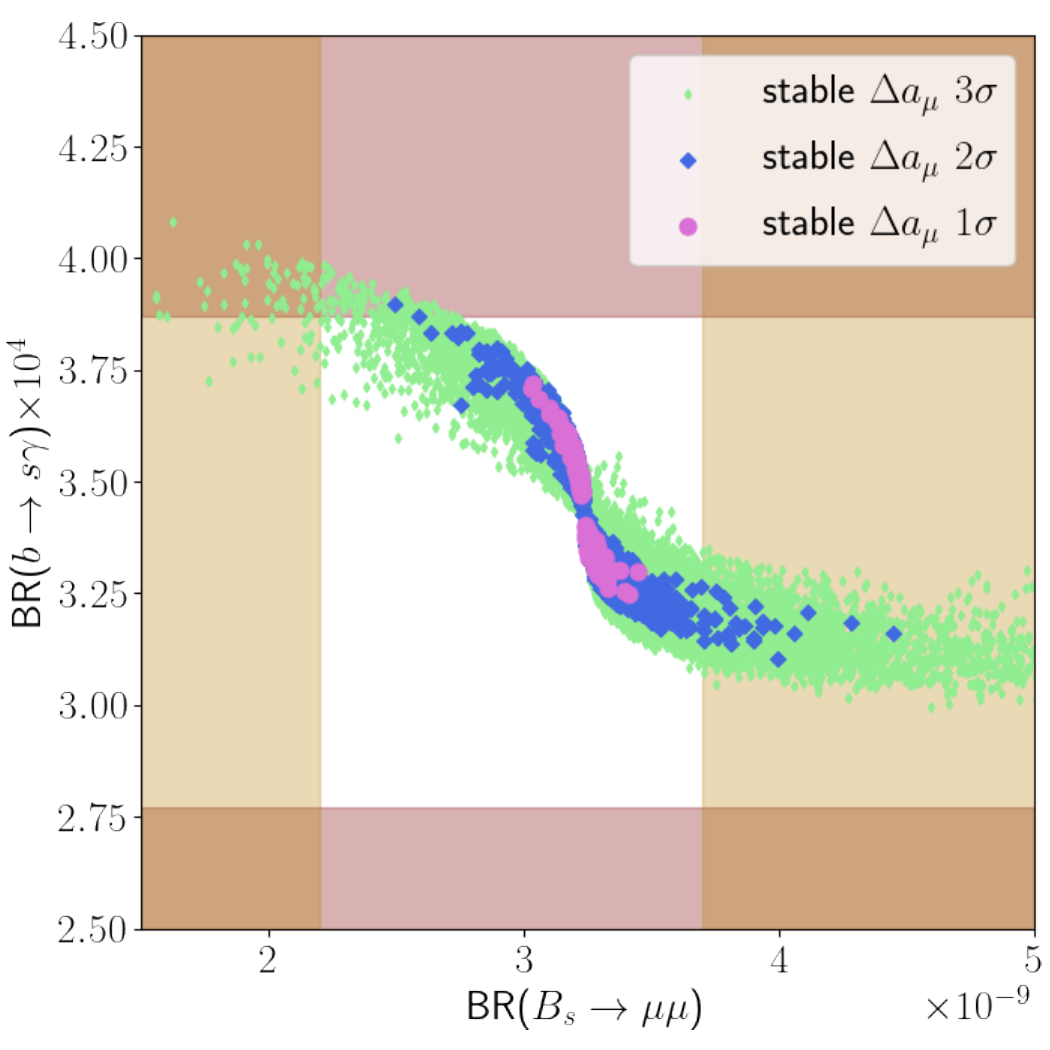}
  \caption{\label{fig-scat_bsgm}
    Scattering plot on ($\br{B_s}{\mu\mu}$, $\br{b}{s\gamma}$).
    The color coding of the points is the same as in the previous
    figures.
    The colored regions deviate from the experimental central values
    by more than $2\sigma$.}
\end{figure}

We now discuss our analysis results.
In Fig.~\ref{fig-scat_tbalp}, we display the parameter points of $\alpha$
and $\tan\beta$, which can explain the muon $g-2$ anomaly within
$3\sigma$, on the left and right panels, respectively.
The results show that the muon $g-2$ prefers the negative values of
$\alpha$, where bino is light: $M_1 \simeq 100$--$200$~GeV.
In particular, the $\Delta a_\mu$ value is the most sizable when
$\alpha \simeq -0.8$ or $-1.6$.
In the former case, $M_1$ is positive, whereas in the latter case, it
is negative as can be seen in Fig.~\ref{fig:gaugino_mass}.
There is a gap between the two cases, where $M_1$ becomes very
small. In the gap, either the $S = {\rm Tr}[Y_i m_{\phi_i}^2]$ or
the $A_\tau$ terms can drive the right-handed stau tachyonic in the
RG running.
The right panel of Fig.~\ref{fig-scat_tbalp} shows that a wide range
of $\tan\beta$ can be compatible with the measured $\Delta
a_\mu$.
However, when $\tan\beta$ is large ($\gtrsim 30$), a deeper
charge-breaking vacuum can be induced even if $\Delta a_\mu$ will be
enhanced by $\tan\beta$ as shown in Eq.~(\ref{eq:a_mu_BLR}).

The SUSY contribution to the muon $g-2$ in mirage mediation
has been studied in Ref~\cite{Cho:2011rk}.
%
Compared to the previous study,
we find that a larger mass hierarchy between the wino/bino and the gluino
is required in order to enhance $\Delta a_\mu$ while avoiding the experimental constraints,
which is for $\alpha$ between about $-3$ and $-0.5$,
as can be seen in Fig.~\ref{fig-scat_tbalp}.
Here, the constraints include the lower bound on the gluino mass from the LHC searches
and the mass of the SM-like Higgs boson.
%
Our analysis shows that the gluino has mass,   $m_{\tilde g} \gtrsim 2.5$~TeV,   in the
parameter region compatible with the muon $g-2$ anomaly.
Another consequence of
$\alpha$ in the indicated region
is that the heavy gluinos drive the up-type
Higgs soft mass squared, $m^2_{H_u}$,
to negative and large in magnitude via RG evolution.
This implies heavy Higgsinos because the EW symmetry breaking requires
\begin{equation}
  |\mu|^2 \approx - m_{H_u}^2 - \frac{1}{2} m_Z^2,
\end{equation}
for moderate to large $\tan\beta$.
It turns out that $|\mu| \gtrsim 2$~TeV in the parameter region for the
muon $g-2$.
Consequently,
  for $\alpha$ in the indicated region,
the SUSY contributions are dominated by the BLR loop diagram, which is
approximately proportional to the Higgsino mass.
The BLR contribution can be enhanced further in the presence of light
smuons below 1~TeV.
We show the sparticle masses in the next section, and benchmark points
are given in Appendix~\ref{sec:benchmark}.

As described in Sec.~\ref{sec:kklt}, the modular weights $c_i$ are
rational numbers in the KKLT setup.
In our analysis, we have taken them to be real positive or
negative numbers to find the viable ranges of $c_i$ that can explain
the muon $g-2$ anomaly.
The result of our parameter scan is shown in Fig.~\ref{fig-scat_cs}.
We find that
the favored regions are $0 \lesssim c_{10} \lesssim 0.5$,
and $-1 \lesssim c_5 \lesssim 0$,
$2 \lesssim c_{H_u} \lesssim 4$, and $-3\lesssim c_{H_d} \lesssim 0$,
which are compatible with the measured $\Delta a_\mu$ value within $2\sigma$.
Among them, $c_5$ and $c_{10}$ are important for
having sizable $\Delta a_\mu$ by the light smuons.
Even if the modulus-mediated contributions to the squarks are small,
the squarks can be heavy due to the RG effects of heavy gluinos.
The $c_{10}$ parameter is related to the mass and the mixing of top
squarks.
In our analysis result, it is mostly positive for achieving $m_h
\simeq 125$~GeV without having too high SUSY-breaking scale.
The $c_{H_u}$ parameter also plays an important role to have a stable
vacuum as it affects the $\mu$ value through the condition of EW
symmetry breaking.
The negative values of $c_{H_d}$ is favored because it lifts up the
slepton masses
through mixed anomaly-modulus mediation and RG running effects.
In Appendix~\ref{sec:benchmark}, we list benchmark sparticle mass
spectra.

Before closing this section, let us discuss the constraints from
flavor-violating processes.
We have calculated the flavor-violating observables by using
\texttt{SuperIso}~\cite{Mahmoudi:2007vz, Mahmoudi:2008tp, Mahmoudi:2009zz}.
Figure~\ref{fig-scat_bsgm} shows the scattering plot of
$\br{B_s}{\mu\mu}$ and $\br{b}{s\gamma}$.
The measured value
of $\br{b}{s\gamma}$ is $(3.32\pm 0.15)\times 10^{-4}$~\cite{Amhis:2019ckw},
and the SM prediction is $(3.36\pm 0.23)\times 10^{-4}$~\cite{Misiak:2015xwa}.
We refer to Ref.\cite{Altmannshofer:2021qrr}
for the combined measurements of $\br{B_s}{\mu\mu}$,
$(2.93\pm0.35)\times 10^{-9}$,
and the SM prediction, $(3.67\pm 0.15)\times 10^{-9}$.
The colored regions in Fig.~\ref{fig-scat_bsgm} are outside the
$2\sigma$ ranges from the experimental central values.
The uncertainties have been obtained by quadrature sums of the SM and
the experimental errors.
The points with larger $\Delta a_\mu$ tend to have smaller SUSY
contributions to the flavor-violating processes because the dangerous
points with light sleptons and large $\tan\beta$ have already been
excluded by requiring the vacuum stability condition.
We have also checked that all the other flavor-violating observables
calculated with \texttt{SuperIso} are consistent with the SM
predictions within current uncertainties.


\section{\label{sec:lhc}Collider signatures and LHC constraints}

\begin{figure}[ph!]
  \begin{center}
    \includegraphics[width=0.45\textwidth]{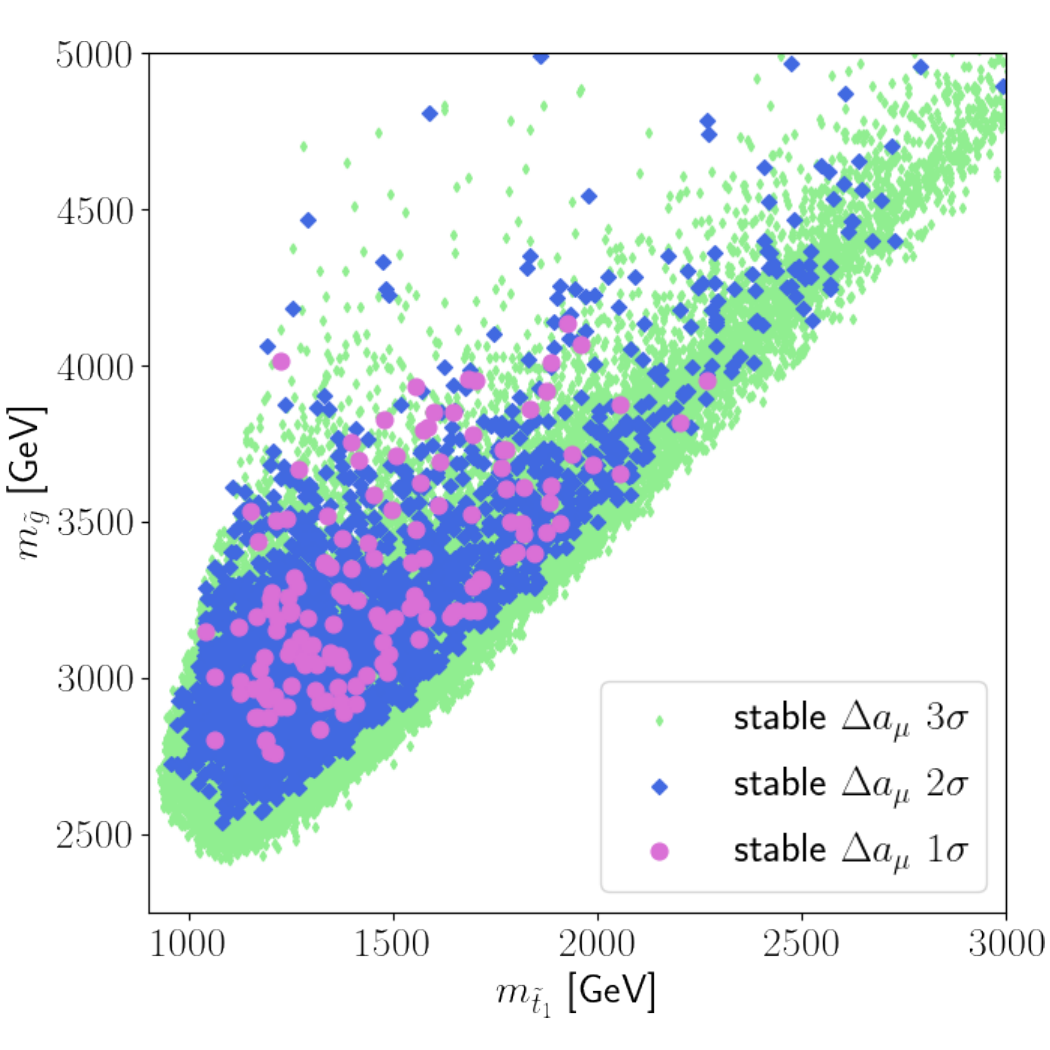}
    \includegraphics[width=0.45\textwidth]{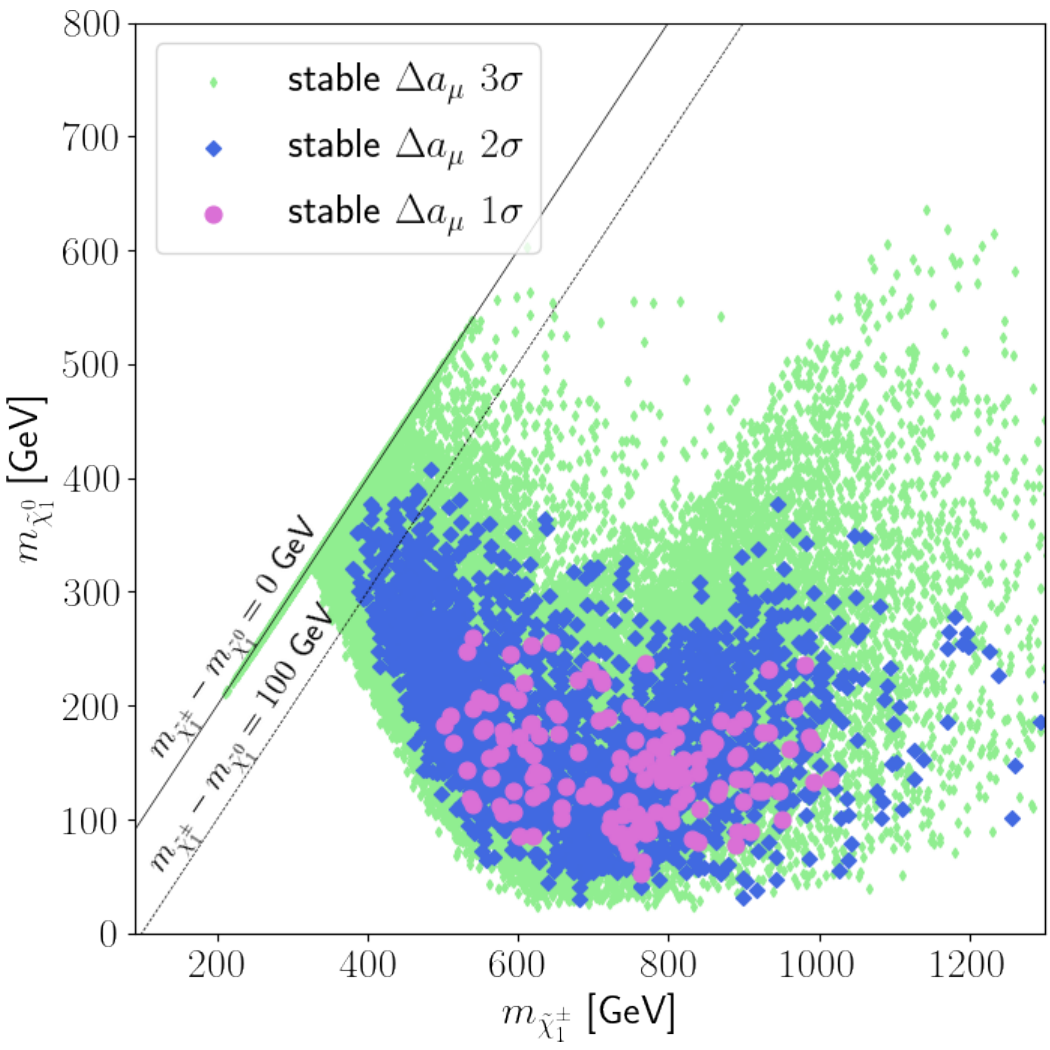} \\
    \includegraphics[width=0.45\textwidth]{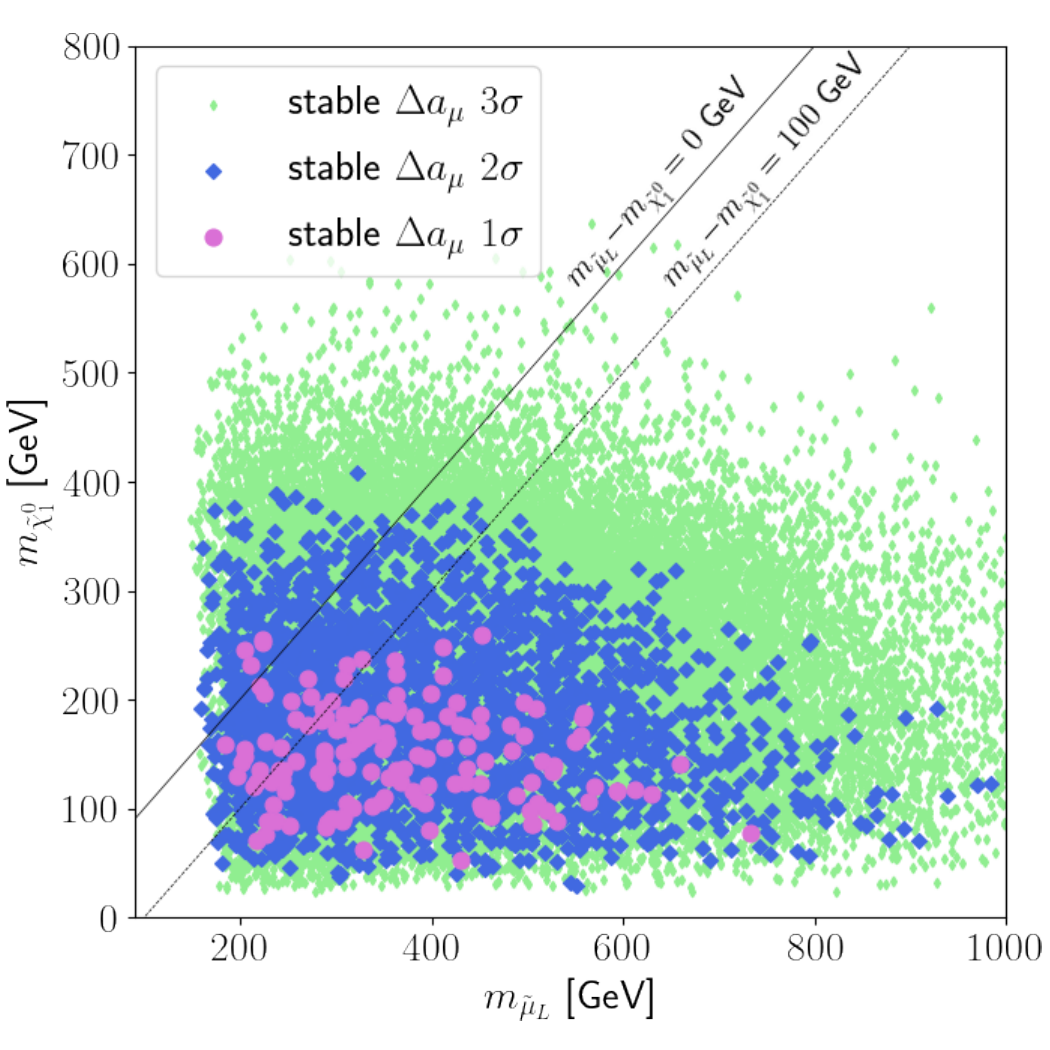}
    \includegraphics[width=0.45\textwidth]{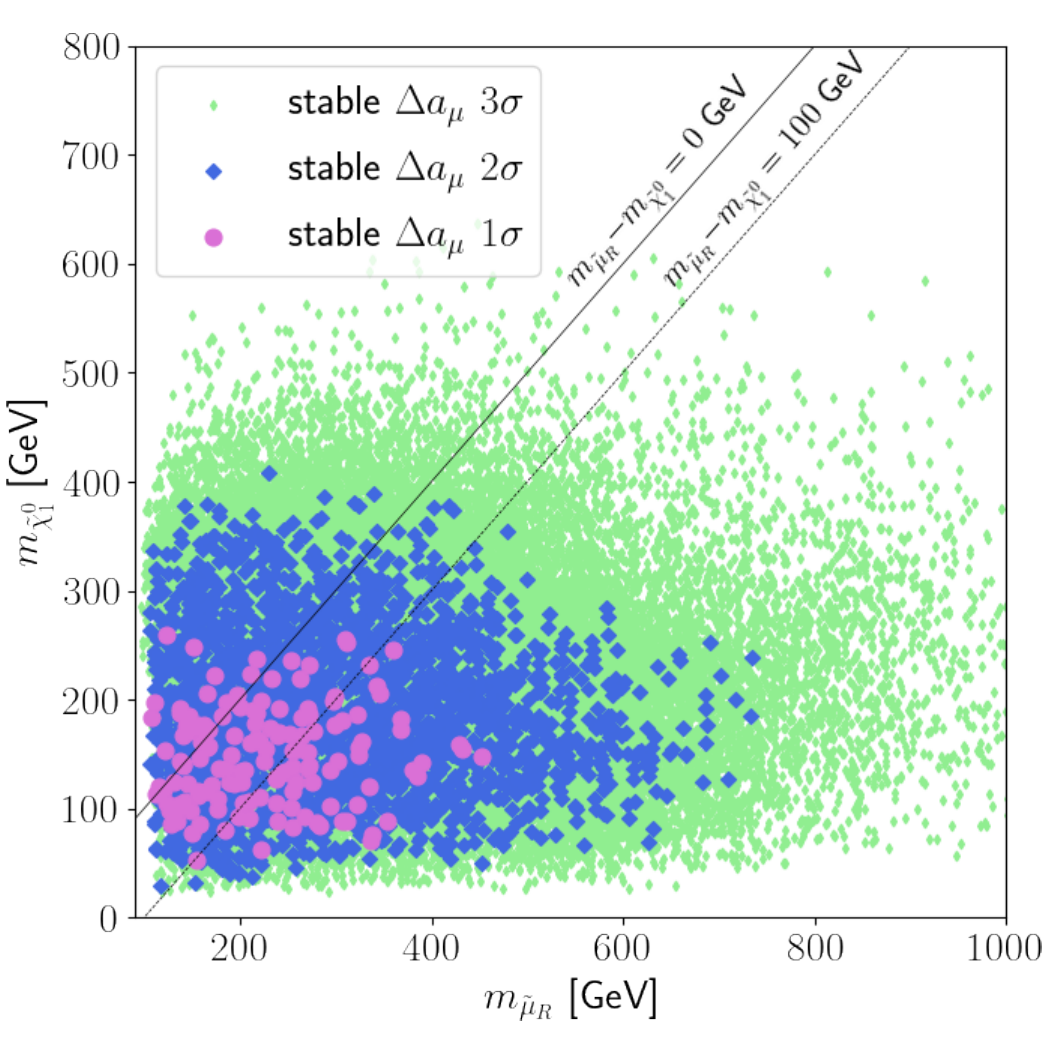} \\
    \includegraphics[width=0.45\textwidth]{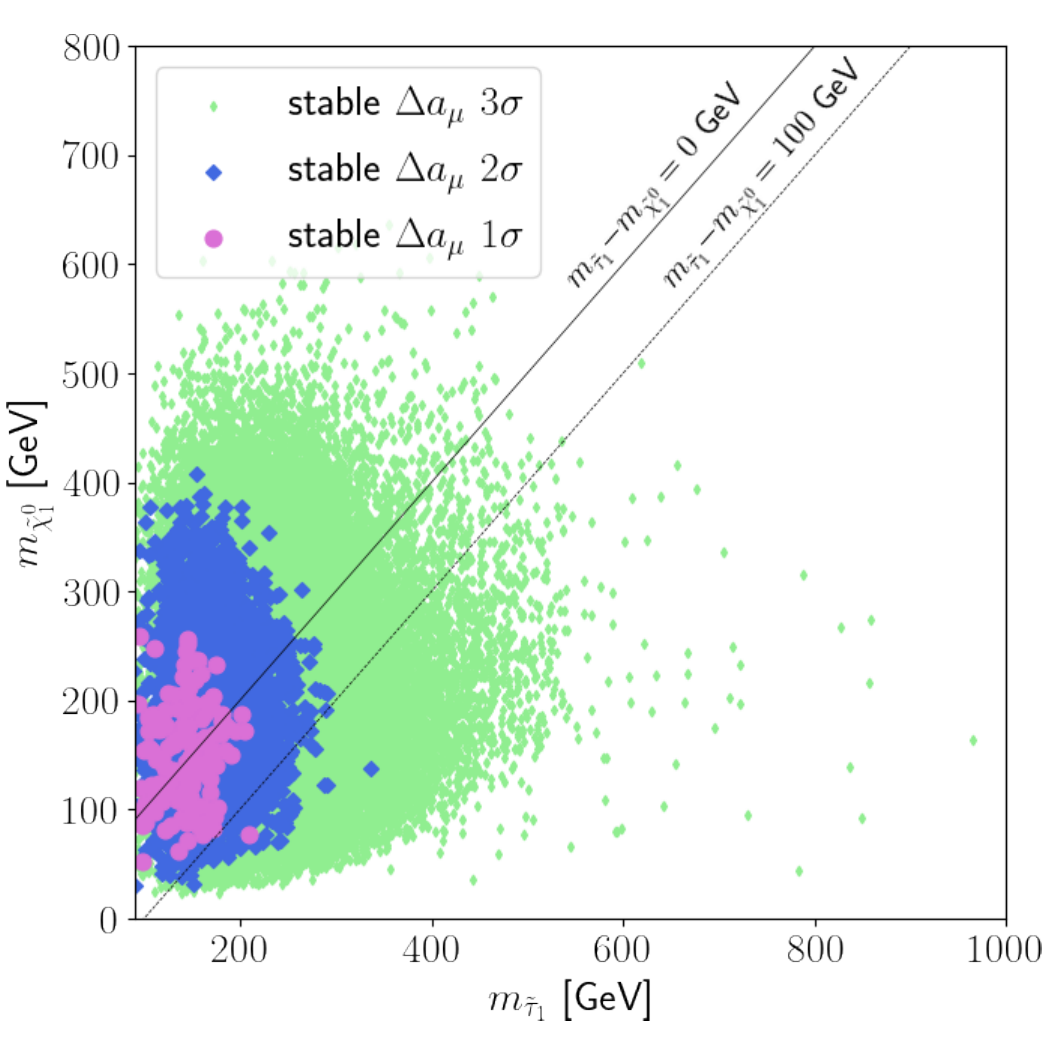}
    \includegraphics[width=0.45\textwidth]{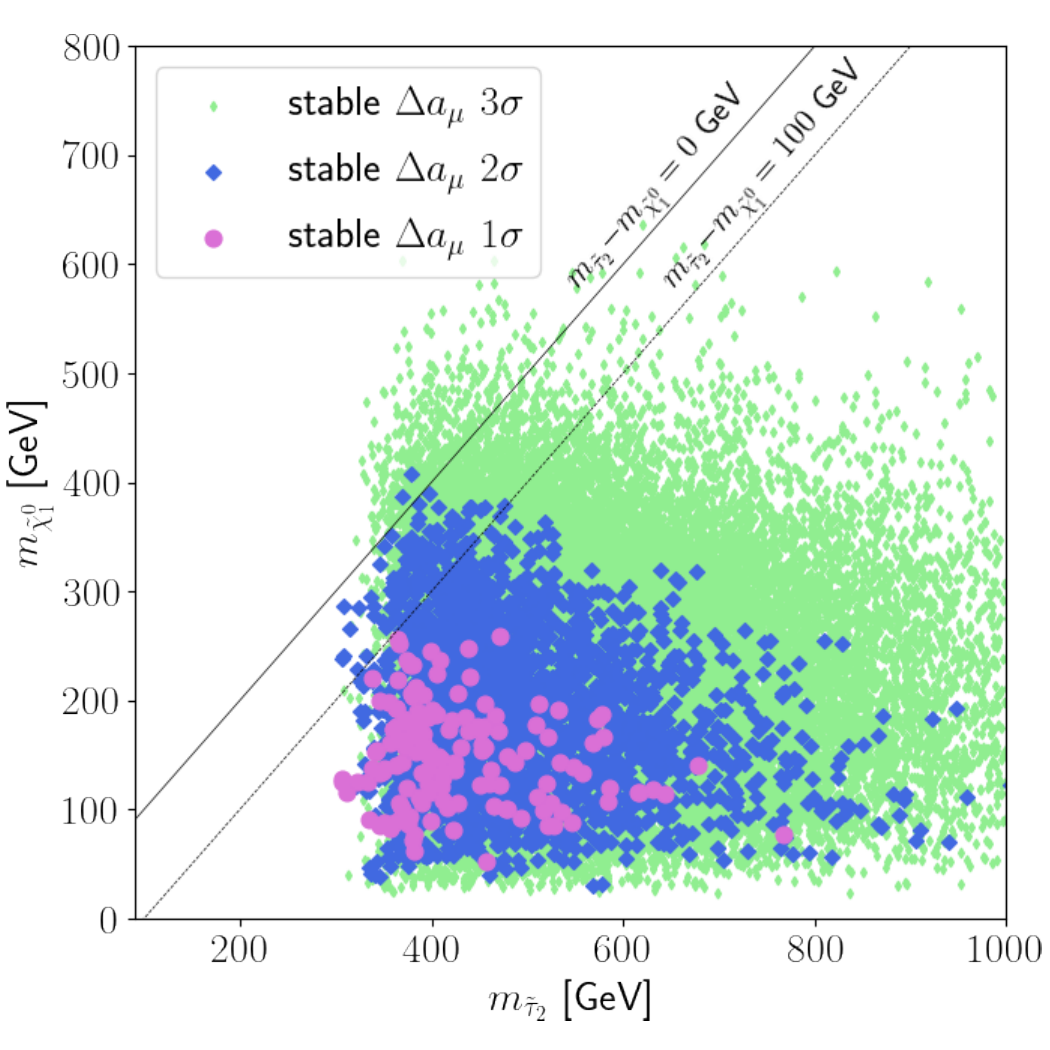}
  \end{center}
  \caption{\label{fig-mass_spectrum}
    The masses of $(\tilde t_1$, $\tilde g)$ (upper left),
    $(\tilde\chi_1^\pm$, $\tilde \chi_1^0)$ (upper right),
    $(\tilde \mu_L$, $\tilde \chi_1^0)$ (middle left),
    $(\tilde \mu_R$, $\tilde \chi_1^0)$ (middle right),
    $(\tilde\tau_1$, $\tilde \chi_1^0)$ (lower left),
    and
    $(\tilde\tau_2$, $\tilde \chi_1^0)$ (lower right)
    in GeV
    for the parameter points
    compatible with the muon $g-2$ anomaly. The color scheme is the same
    as in Fig.~\ref{fig-scat_cs}.}
\end{figure}

\noindent
In this section, we discuss viable phenomenological scenarios of mixed
modulus-anomaly mediation motivated by the muon $g-2$ anomaly and the
relevant experimental constraints.
Among the sparticles,
in general,
the colored sparticles receive the most severe constraints from the
SUSY searches at hadron colliders.
The latest LHC Run 2 analysis results of the ATLAS~\cite{Aad:2020aze}
and CMS~\cite{Sirunyan:2019xwh} collaborations have excluded the
gluino mass below 2.3~TeV.
In our study, the $M_0$ value has been fixed by requiring the Higgs
mass to be compatible with the measured SM-like Higgs mass, given the
other model parameters.
It results in a large value of $M_0$ that leads to heavy gluinos.
Furthermore,   as seen in Sec.~\ref{sec:kklt},
a negative $\alpha$
can raise
the gluino mass up to multi-TeV scales
while
leaving the EWinos
around the weak scale.
The upper left panel of Fig.~\ref{fig-mass_spectrum} shows
that $m_{\tilde g} \gtrsim 2.5$~TeV in the parameter space that can
explain the muon $g-2$ anomaly within $2\sigma$.
On the other hand,
the sfermion masses
%
have a strong dependence on  $M_0$ and $c_i$.
In the parameter space for the muon $g-2$, the masses of the lighter
stop are close to or slightly above the current lower limit, which is
$m_{\tilde t_1} \gtrsim 1.2$~TeV~\cite{Aad:2020aob, Sirunyan:2021mrs}.
Consequently, we expect that the parameter points having stop
masses around or above 1~TeV will be tested by searches at the future
LHC Run 3 and the High-Luminosity LHC\@.
The other squark masses such as the lighter sbottom are well above the
current experimental bounds.
In Appendix~\ref{sec:benchmark}, we present the benchmark sparticle mass
spectra.

Contrary to the colored sparticles, the lighter chargino and
neutralinos, as well as the sleptons, have masses around the weak scale
to explain the muon $g-2$ anomaly.
Therefore, the search results on the direct productions of the
neutralinos/charginos and the sleptons can impose more serious limits
on the parameter space than the experimental bounds discussed above.
We exhibit the lighter chargino and slepton masses in
Fig.~\ref{fig-mass_spectrum}.
The lighter chargino is dominantly wino-like because the Higgsino is much
heavier, $\abs{\mu}\gtrsim 2$~TeV,
and hence the second lightest neutralino has degenerate mass with the chargino.
The left-handed sleptons tend to be heavier than the right-handed ones
due to the RG effects from relatively large wino mass.
The selectrons are nearly mass degenerate with the smuons,
whereas the stau can have different masses than the other
sleptons due to the left-right mixing terms and the RG effects.

Even though the muon $g-2$ anomaly hints at the existence of bino and
smuons around the weak scale, it can lead to various phenomenological
scenarios depending on the interactions and the mass spectrum
of the sparticles in the low-energy scale.
Classifying them by the property of the LOSP, we consider three
phenomenological scenarios:
\renewcommand*\labelenumi{(\theenumi)}
\begin{enumerate}
 \item {The neutralino is the LOSP and is stable.}
 \item {The charged slepton is the LOSP and is metastable.}
 \item {The LOSP is unstable due to the RPV.}
\end{enumerate}
Here, being stable means that the particle does not necessarily be
completely stable: it does not decay inside detectors at collider
experiments.

\subsection{Stable  neutralino LOSP}

In the first scenario, the LOSP is the lightest neutralino
$\tilde\chi_1^0$.
Because the Higgsino is very heavy, $|\mu| \gtrsim 2$~TeV, in the
parameter space for the muon $g-2$, $\tilde\chi_1^0$ is dominantly
bino-like or an admixture of wino and bino.
In this scenario, the most stringent limits come from the search
results on the direct productions of the neutralino-chargino
$\tilde\chi_2^0 \tilde\chi_1^\pm$ and the slepton pair $\tilde \ell
\tilde \ell$ at the LHC~\cite{Aad:2019vnb, CMS-PAS-SUS-19-012, Sirunyan:2020eab}.
Assuming mass-degenerate left-handed (right-handed) light flavor sleptons,
the search results for the slepton pair productions
have set the lower limit for the slepton mass
to be $m_{\tilde \ell_L} > 650$~GeV ($m_{\tilde \ell_R} > 500$~GeV)
for $m_{\tilde \chi_1^0}$ being up to 400~(200)~GeV~\cite{Aad:2019vnb}.
Searches for stau pair productions can also give constraints
because the stau is often the next-to-lightest SUSY particle
in a portion of the parameter space with the neutralino LOSP that can
explain the muon $g-2$ anomaly in our setup.
However, the current limits for the staus decaying into the
neutralino LOSP are not very stringent,   compared to those for  the
charginos and the other sleptons~\cite{CMS:2019eln, Aad:2019byo}.
A way out of the LHC constraints is to have sleptons 
mass-degenerate with the neutralino LOSP,
$m_{\tilde \ell}-m_{\tilde \chi_1^0} \lesssim 80$~GeV.
The limits for such degenerate spectrum is quite restricted~\cite{Aad:2019qnd}.

The search results on the neutralino-chargino production at the LHC set
the stringent limits on the chargino mass~\cite{Aad:2019vnb, CMS-PAS-SUS-19-012}.
If the mass gaps among the sleptons and the winos are
sufficiently large, decays to all lepton flavors occur with almost
equal probability, i.e., flavor-democratic decays.
On the other hand, the wino-like states will dominantly decay into a
stau and a tau (neutrino) if the other decay modes are kinematically
forbidden.
In the flavor-democratic case,
the CMS analysis result for the integrated luminosity of 137~fb$^{-1}$
has excluded the wino-like chargino mass up to about 1.3~TeV when
$m_{\tilde{\chi}^0_1} \lesssim 800$~GeV~\cite{CMS-PAS-SUS-19-012}.
We see from Fig.~\ref{fig-mass_spectrum} that, in the
flavor-democratic case, all the parameter points are excluded by the
CMS search result except for the mass-degenerate region.
However, in the mass-degenerate region with $\alpha \lesssim -2.5$,
the bino becomes heavy, and thus the sleptons must be light to explain
$\Delta a_\mu$.
Consequently, the flavor-democratic decays of the wino-like states are
not achievable in the parameter region compatible with the muon $g-2$
anomaly.
Meanwhile,
if the wino-like states dominantly decay to a stau and a tau (neutrino),
the lower limit for the chargino mass is about
800~GeV for $m_{\tilde{\chi}^0_1} \lesssim 100$ GeV~\cite{CMS-PAS-SUS-19-012},
which is much weaker than that in the flavor-democratic case.
This can be realized
if the winos are lighter than the left-handed selectron and smuon.

We conclude that, in the case of neutralino LOSP, the current LHC
limits can be satisfied if the right-handed sleptons are nearly
degenerate with the neutralino LOSP, and the left-handed selectron and
smuon are sufficiently heavy so that the wino-like states dominantly
decay to the stau:\footnote{
  Here, we have used the ``$\lessapprox$'' symbol to indicate that the
  two particles are close in mass while evading the LHC limits by
  following the notation in Ref.~\cite{Athron:2021iuf}.
}
\begin{equation}
\label{eq-noLHCspec}
  m_{\tilde{\chi}^0_1} \lessapprox m_{\tilde{\ell}_R},
  \quad
  m_{\tilde{\chi}^0_1} \lesssim
  m_{\tilde \tau_1} \lesssim m_{\tilde\chi_1^\pm, \tilde\chi_2^0}
  \lesssim m_{\tilde e_L, \, \tilde \mu_L}, \quad
  m_{\tilde e_L, \, \tilde \mu_L} > 650~{\rm GeV} .
\end{equation}
In addition to the above mass hierarchy, one should ensure that the
right-handed slepton is heavier than about 100~GeV to avoid the lower
limit from LEP on the slepton masses.
The limit also applies indirectly to $m_{\tilde\chi_1^0}$ in the case
where the bino-like neutralino LOSP is degenerate in mass with the
slepton.
Furthermore, if $m_{\tilde\chi_1^0} \lesssim 100$~GeV, the lower
limit from CMS on the chargino mass, $m_{\tilde\chi_1^\pm} \gtrsim
800$~GeV, should also be taken into account.
The benchmark point A shown in Appendix~\ref{sec:benchmark}
corresponds to this scenario.
It is interesting that the neutralino LOSP
could serve as a good dark matter (DM) candidate through slepton co-annihilations
due to the mass degeneracy.\footnote{
The bino-like neutralino LOSP around the weak scale may be
overproduced via late-time decays of a modulus~\cite{Endo:2006zj}
unless the modulus is located quite close to the potential minimum
after the primordial inflation.
This fact also motivates us to consider the axino as the lightest SUSY
particle (LSP) or RPV scenarios, which will be discussed in the
following subsections.
}

%

\subsection{Metastable slepton LOSP}

If the slepton is the LOSP,
there should be lighter sparticle than it or R-parity should be violated
so that the slepton LOSP can decay. The latter scenario will be
discussed in the next subsection. In the former case, a scenario worth
considering is a PQ symmetric extension where the axion solves the
strong CP problem, and the axino $\tilde a$  contributes to the DM~\cite{Covi:1999ty}.
For instance,
if the saxion is radiatively stabilized~\cite{Nakamura:2008ey,Choi:2009qd},
the axino  naturally becomes the LSP because its mass is one-loop suppressed
compared to other sparticle masses.
%
The scenario is noteworthy because, in a majority of the
parameter space compatible with the muon $g-2$ anomaly, we find that
a slepton is lighter than the lightest neutralino:
the stau is the LOSP in more than half of the
parameter space and the selectron or the smuon is the LOSP in many
other parameter points.
Then, the slepton LOSP will mainly undergo the two-body decay,
$\tilde \ell \to \ell \tilde a$.

In this scenario, the slepton LOSP becomes a heavy stable
charged particle (HSCP), mostly decaying outside the detector, because
it can have a lifetime longer than $10^{4}$~ns to a few hundred
seconds, depending on the axion decay constant and the masses of
the involved sparticle masses~\cite{Brandenburg:2005he}.
Due to the long lifetime, the scenario is not constrained by
the search results for displaced leptons because it is sensitive to
the particles with lifetime shorter than 1~ns~\cite{Aad:2020bay}.
At LEP2, the null detection of the HSCPs set the lower mass limit of
about 100 GeV~\cite{Abbiendi:2003yd}.
In recent years, the constraint has been updated further by the
searches for HSCPs at the LHC\@.
In particular, the CMS collaboration performed model-independent
analyses for various possible HSCPs and excluded stau masses
below 360~GeV~\cite{CMS-PAS-EXO-16-036}.
In the scenario with the axino LSP,
the CMS limit can impose a serious impact on our analysis result
because the lighter stau is lighter than about 350~GeV in the
parameter points with the stau LOSP for the muon $g-2$ anomaly within
$3\sigma$, as can be seen in the lower left panel of Fig.~\ref{fig-mass_spectrum}.

\subsection{Unstable LOSP}

The LOSP, either neutralino or slepton,
decays to SM particles if the RPV interactions are allowed.
In the RPV scenario, the axion can serve as a candidate for the DM~\cite{Baer:2014eja}.
The relevant RPV terms in the superpotential are given as follows:
\begin{equation}
  W \supset \frac{1}{2} \lambda_{ijk} L_i L_j \bar E_k +
  \lambda_{ijk}^\prime L_i Q_j \bar D_k + \frac{1}{2}
  \lambda_{ijk}^{\prime \prime} \bar U_i \bar D_j \bar D_k ,
\end{equation}
where $i$, $j$, $k$ are flavor indices. The $\lambda_{ijk}$ and
$\lambda_{ijk}^\prime$ terms violate lepton number while the
$\lambda_{ijk}^{\prime \prime}$ terms violate baryon number.
Either of the lepton or baryon number conservation should hold with good accuracy
to avoid too fast proton decay.

If the terms with the $\lambda_{ijk}$ couplings are dominant among
the others, the slepton LOSP will mainly decay into a charged lepton and a
neutrino, $\tilde \ell \to \ell_j \nu_{\ell_k}$.
The slepton pair
production then gives rise to the signature of $2\ell+E_T^\mathrm{miss}$.
The signature is similar to that of the R-parity conserving case,
and it receives the bounds from the aforementioned SUSY searches for
multi-lepton final states.
Recasting of the LHC search results has revealed that the
lower limit of the stau LOSP is about 225~GeV in the case where the
stau is right-handed~\cite{Dreiner:2020lbz}.
A large portion of our parameter space could be excluded by the limit.
Meanwhile, if either $\lambda_{ijk}^\prime$ or $\lambda_{ijk}^{\prime
  \prime}$ term is dominant, the final states of the slepton decays
are quite different. In the case where the $\lambda_{ijk}^{\prime}$
operators are dominant, the slepton LOSP can decay into
the final state of two leptons $+$ two quark-jets via
four-body processes. For example, the stau LOSP can decay as
\begin{equation}
  \tilde \tau_1 \to \tau + \tilde\chi_1^{0 \ast} \to \tau + \mu\,
  u \, \bar d
\end{equation}
via the $\lambda_{211}^{\prime}$ operator.
See Refs.~\cite{Desch:2010gi, Dercks:2017lfq}
for a list of possible LHC signatures.
The decay length of the stau LOSP can be $\mathcal{O}(10^{-6})$~m
for $\lambda^\prime \simeq 10^{-3}$ and $m_{\tilde \tau_1} \simeq
m_{\tilde\chi_1^0} \simeq 100$~GeV,
resulting in displaced vertices~\cite{Allanach:2003eb}.
As the gluino and squark masses are beyond the current experimental
limits, the slepton LOSP could be produced via the neutralino-chargino or
the direct slepton pair processes.
The signatures have not yet been covered by the LHC searches so far.
Therefore, we conclude that the slepton LOSP scenario with RPV would
be viable unless the $\lambda_{ijk}$ operator is the dominant RPV
interaction.

In the case of neutralino LOSP with RPV, we can reach a similar conclusion.
If R-parity is violated dominantly by the $\lambda_{ijk}$
coupling, there are strong constraints due to the signatures of
high-multiplicity leptons~\cite{Aad:2021qct,Aad:2020cqu}.
For instance, the limits for sleptons and charginos are
about 800 and 1000~GeV, respectively, in the scenario of nonzero
$\lambda_{i33}$~\cite{Aad:2021qct}.
In the other cases where either $\lambda_{ijk}^\prime$ or
$\lambda_{ijk}^{\prime \prime}$ is dominant, the limits are much
weaker or absent.

\section{\label{sec:summary}Summary}

\noindent
Since the new measurement of the muon $g-2$ at the Fermilab
experiment,
physicists have regained attention on the existence of new
physics in the lepton sector.
If new physics responsible for the muon $g-2$ anomaly is
supersymmetric, one should consider how to obtain light
EWinos and sleptons in a systematic way.
Combined with the gauge coupling unification,
the gaugino masses exhibit a robust pattern controlled by a single parameter $\alpha$
that  represents the size of anomaly mediation.
The EWinos can be much lighter than the gluino if $\alpha$ is negative and of order unity,
as is required to explain the muon $g-2$ anomaly while avoiding experimental constraints.
The KKLT provides a natural and interesting framework for such mixed mediation,  where
the pattern of gaugino masses is determined by $\alpha$,  while that of sfermion masses
depends on how the corresponding matter field couples to the string moduli sector.

We have performed a numerical analysis to explore the parameter space
of mixed modulus-anomaly mediation
realized in the generalized KKLT setup
and identified the parameter region
compatible with the muon $g-2$ anomaly.
To have light EWinos,
it is essential to construct a setup of KKLT moduli stabilization
yielding a negative $\alpha$.
As a byproduct,  it can make
the gluino
heavier than a few TeV, thus we can easily evade the lower limit of
gluino at the LHC\@.
On the other hand, due to light sleptons, imposing the condition of
vacuum stability of the scalar potential is crucial,    and it excludes
the parameter space of large $\tan\beta \gtrsim 30$.

In the viable parameter region,  we find that the LOSP can be either
bino-like neutralino or slepton.
However, in the case of the neutralino LOSP,
the slepton and chargino-neutralino searches at the LHC
exclude
a vast parameter space of the R-parity conservation.
%
The current LHC limits can be satisfied only when the mass spectrum
of Eq.~\eqref{eq-noLHCspec} is realized.
In most cases, the wino cannot be sufficiently heavy or degenerate with the bino
as far as the sleptons are sufficiently light due to the gaugino mass relations
predicted in the mixed modulus-anomaly mediation.
To avoid this difficulty, one may consider a more general case with
$c_i \neq a_i$. Another way is to add gauge-mediated contributions so
that the deflection of sparticle masses occurs at the gauge-messenger
scale~\cite{Everett:2008qy,Everett:2008ey}.
Meanwhile,  when a slepton is  lighter than the neutralinos,  we should consider
alternative scenarios such as axino LSP or RPV interactions.
In the former case with axino LSP, the lightest slepton becomes
long-lived and will decay outside the detector.
The recent CMS result on long-lived charged particles has excluded
such possibility.
On the other side, the RPV interactions with either lepton or baryon
number violation can be a viable option because of unexplored
signatures with the final states of multi-jets and -leptons with small
or no missing energy at the LHC\@.

\section*{Acknowledgments}

\noindent
This work was supported by the National Research Foundation of Korea (NRF)
grant funded by the Korean government NRF-2018R1C1B6006061 (K.S.J.),
the Institute for Basic Science (IBS) under the project code
IBS-R018-D1 (J.K. and C.B.P.), and
the Grant-in-Aid for Scientific Research from the Ministry of
Education, Science, Sports and Culture (MEXT), Japan No.\ 18K13534
(J.K.).


\appendix
\section{\label{sec:benchmark}Benchmark sparticle mass spectrum}

The sparticle mass spectra of our benchmark points in mixed
modulus-anomaly mediation are shown in Table~\ref{tab-spectrum}.
At points A and B, the LOSP is bino-like, while it is the lightest
stau at points C and D.
The sign of bino mass $M_1$ is taken to be positive at points A and C,
while it is negative at points B and D.
At points A and B, the right-handed sleptons are degenerate with the
lightest neutralino, and the left-handed ones are sufficiently heavy
so that the current limits can be evaded.
Point A, which realizes the mass spectra of Eq.~\eqref{eq-noLHCspec},
can be safe from the constraint from the latest CMS search results on the
chargino-neutralino productions because the stau is lighter than the
wino-like states, while the other left-handed sleptons are heavier.
In this case, $\tilde \chi_1^\pm$ and $\tilde \chi_2^0$ dominantly
decay into the stau and the tau (neutrino).
Meanwhile, point B is excluded by the CMS search result because
the wino-like states decay into the sleptons with nearly equal
branching fraction to each flavor.
Points C and D could be excluded by the HSCP searches at the LHC if
the stau is metastable.
However, if R-parity is violated mainly by $\lambda^\prime$ or
$\lambda^{\prime\prime}$, all the points are still viable and can be
searched at the LHC or future colliders through the final states of
multi-leptons and jets.

\begin{table}[t]
\centering
\caption{\label{tab-spectrum}
  Benchmark sparticle mass spectra of mixed modulus-anomaly mediation
for the muon $g-2$ anomaly.
}
\footnotesize
\begin{tabular}[t]{c|cccc} \hline
 & A & B & C & D \\ \hline\hline
 $\tan\beta$ & 30.41 & 9.955 & 23.63 & 12.9 \\
 $\mathrm{sgn}(\mu)$ & 1 & $-1$ & 1 & $-1$ \\
$M_0$ & 1038 & 1062 & 1250 & 1344 \\
$\alpha$ & $-0.7734$ & $-1.51$ & $-0.5258$ & $-1.745$ \\ \hline
$c_Q$ & 0.06998 & 0.1594 & 0.1025 & 0.05333 \\
$c_U$ & 0.06998 & 0.1594 & 0.1025 & 0.05333 \\
$c_D$ & $-0.1951$ & $-0.8861$ & $-0.5709$ & $-0.8574$ \\
$c_L$ & $-0.1951$ & $-0.8861$ & $-0.5709$ & $-0.8574$ \\
$c_E$ & 0.06998 & 0.1594 & 0.1025 & 0.05333 \\
$c_{H_u}$ & 2.892 & 3.829 & 2.312 & 2.373 \\
$c_{H_d}$ & $-1.665$ & $-0.9771$ & $-0.9358$ & $-0.899$ \\  \hline \hline
$m_h$ & 125.1 & 125.1 & 125.1 & 125.1 \\
$m_A$ & 1735 & 3021 & 1952 & 3451 \\ \hline
$m_{\tilde{g}}$ & 2864 & 3507 & 3175 & 4584 \\
$m_{\tilde{\chi}_1^0}$ & 157.9 & 120.1 & 301.7 & 262 \\
$m_{\tilde{\chi}_2^0}$ & 767.8 & 704.4 & 950.1 & 861.2 \\
$m_{\tilde{\chi}_1^\pm}$ & 768 & 704.5 & 950.2 & 861.4 \\
$m_{\tilde{\chi}_3^0}$ & 2062 & 3021 & 2111 & 3453 \\
$m_{\tilde{\chi}_4^0}$ & 2064 & 3021 & 2113 & 3453 \\
$m_{\tilde{\chi}_2^\pm}$ & 2065 & 3022 & 2114 & 3454 \\ \hline
$m_{\tilde{b}_1}$ & 2133 & 2579 & 2427 & 3649 \\
$m_{\tilde{b}_2}$ & 2487 & 3040 & 2624 & 3990 \\
$m_{\tilde{t}_1}$ & 1304 & 1237 & 1618 & 2600 \\
$m_{\tilde{t}_2}$ & 2160 & 2597 & 2451 & 3661 \\
$m_{\tilde{q}_1}$ & 2698 & 3364 & 2984 & 4376 \\
$m_{\tilde{u}_1}$ & 2639 & 3290 & 2907 & 4262 \\
$m_{\tilde{d}_1}$ & 2526 & 3049 & 2655 & 4014 \\ \hline
$m_{\tilde{\tau}_1}$ & 182.3 & 144.3 & 265.2 & 244.5 \\
$m_{\tilde{\tau}_2}$ & 796.1 & 374.2 & 497 & 449.3 \\
$m_{\tilde{\mu}_L}$ & 775.4 & 213.7 & 367.8 & 346.5 \\
$m_{\tilde{\mu}_R}$ & 163.7 & 334.2 & 416.3 & 350.5 \\
$m_{\tilde{e}_L}$ & 775.3 & 213.6 & 367.8 & 346.4 \\
$m_{\tilde{e}_R}$ & 163.4 & 334.2 & 416.2 & 350.4 \\
$m_{\tilde{\nu}_e}$ & 778.3 & 203.3 & 364.3 & 346.9 \\
$m_{\tilde{\nu}_\mu}$ & 771.1 & 199.1 & 359.2 & 337.4 \\
$m_{\tilde{\nu}_\tau}$ & 771.1 & 199.1 & 359.2 & 337.3 \\  \hline \hline
$\Delta a_\mu$ $\times 10^9$ & 1.643 & 2.275 & 1.442 & 1.433 \\
BR($b \to s \gamma$) $\times 10^4$ & 3.177 & 3.517 & 3.284 & 3.461 \\
BR($B_s\to \mu\mu$) $\times 10^9$ & 3.862 & 3.216 & 3.418 & 3.213 \\ \hline
$\eta_\mu$  & 0.0653 & 0.0562 & 0.0546 & 0.0592 \\
$\eta_\tau$ & 0.8674 & 0.8910 & 0.8683 & 0.9077 \\ \hline
LOSP & ${\tilde{\chi}_1^0}$ & ${\tilde{\chi}_1^0}$ & ${\tilde{\tau}_1}$ & ${\tilde{\tau}_1}$ \\
\hline
\end{tabular}
\end{table}

\clearpage
\bibliography{susy_muon_g-2}

\end{document}